# Closed orbit correction of HIMM synchrotron

WANG Geng(王耿)　　SHI Jian(石健)*　　YANG Jian-Cheng(杨建成)　　XIA Jia-Wen(夏佳文)
RUAN Shuang(阮爽)　　WU Bo(吴波)　　ZHAO He(赵贺)

Institute of Modern Physics, Chinese Academy of Sciences, Lanzhou 730000, China

**Abstract:**　 The correction of closed orbit has great influence on the operation of synchrotron. The design of correction system is one significant component of lattice design. It is suggested to set BPMs at the peaks of betatron oscillation. The correctors need to be located at the positions where β function is large or the sources of large errors. In the simulation of the closed orbit correction of HIMM synchrotron, one important reason affecting the result of horizontal correction is the longitudinal alignment error of dipole magnet. It is advisable to decrease this kind of alignment error while the deflection angle of dipole magnet is large.

## 1　Introduction

Heavy Ion Medical Machine (HIMM) is a new therapy facility dedicated to cancer treatment, designed by Institute of Modern Physics. The main accelerator is a compact synchrotron designed to accelerate $C^{6+}$ to maximum energy of 400 MeV/u, including 8 dipole magnets and 12 quadrupole magnets. The correction system of the closed orbit includes 8 BPMs providing horizontal and vertical closed orbit distortion (COD) measurement, and respectively 6 and 7 correctors in horizontal and vertical directions. In synchrotron, when only linear elements are considered, the factors impacting the closed orbit can be classified by dipole and quadrupole field components. The former which is the main component to be considered arouses the distortion and the latter changes the distribution and oscillation of the COD. COD control is very important for successful operation of synchrotron. The errors resulting in the distortion consist of alignment deviations and field imperfections of the magnets.

## 2　Closed orbit

In synchrotron, assuming that there is a dipole field error at one point causing deflection of θ, and that the transfer matrix of one turn is M, the transverse movement satisfies the equation [1]

$$\begin{pmatrix} x \\ x' \end{pmatrix} = M \begin{pmatrix} x_0 \\ x'_0 \end{pmatrix} + \begin{pmatrix} 0 \\ \theta \end{pmatrix}. \quad (1)$$

Like the solution of nonhomogeneous ordinary differential equation, we only need a special solution. Defining

---
* Corresponding author (email: shijian@impcas.ac.cn)

$$x = x_\beta + x_c, \quad (2)$$

Eq. (1) could be written as

$$\begin{pmatrix} x_\beta + x_c \\ x'_\beta + x'_c \end{pmatrix} = M \begin{pmatrix} x_{\beta 0} + x_{c0} \\ x'_{\beta 0} + x'_{c0} \end{pmatrix} + \begin{pmatrix} 0 \\ \theta \end{pmatrix}, \quad (3)$$

namely

$$\begin{pmatrix} x_\beta \\ x'_\beta \end{pmatrix} + \begin{pmatrix} x_c \\ x'_c \end{pmatrix} = M \begin{pmatrix} x_{\beta 0} \\ x'_{\beta 0} \end{pmatrix} + M \begin{pmatrix} x_{c0} \\ x'_{c0} \end{pmatrix} + \begin{pmatrix} 0 \\ \theta \end{pmatrix}, \quad (4)$$

where $x_\beta$ means β oscillation. Supposing that $x_c$ remains the same, there is relationship

$$\begin{pmatrix} x_c \\ x'_c \end{pmatrix} = M \begin{pmatrix} x_c \\ x'_c \end{pmatrix} + \begin{pmatrix} 0 \\ \theta \end{pmatrix}. \quad (5)$$

Then we will gain a special solution which reflects the closed orbit. Distortions aroused by errors at different points satisfy the superposition principle. Plugging the expression of M and the transfer matrix of two points, we will gain the formula of COD [1]

$$x(s) = \frac{\theta \sqrt{\beta(s)\beta_0}}{2\sin\pi\nu} \cos[|\psi(s) - \psi(0)| - \pi\nu]. \quad (6)$$

It is found that the closed orbit is the reference orbit of β oscillation.

## 3　Correction system

The correction system of an accelerator includes BPMs and correctors. BPM is used to measure the closed orbit. Theoretically, the closed orbit correction should reduce the

CODs at the locations of BPMs to zero. According to formula (6), the distortion is proportional to the sqrt of β value. So it is advisable to set BPMs at the peaks of the betatron oscillation.

The connection between BPMs and correctors is the response matrix defined A, of which the element is $a_{ij}$ having the expression [2]

$$a_{ij} = \frac{\sqrt{\beta_i \beta_j}}{2\sin\pi\nu} \cos[|\Delta\psi| - \pi\nu]. \quad (7)$$

Assuming that there are n BPMs and m correctors, the orbit measured by BPM is $p_i$ and the corrector's strength needed is $b_j$. In an ideal situation, there is relationship [2]

$$\vec{p}_{i,n} = -A_{n\times m} \vec{b}_{j,m}, \quad (8)$$

$$\vec{b}_{j,m} = -(A^T \cdot A)^{-1} A^T \vec{p}_{i,n} = -A^{-1} \vec{p}_{i,n}, \quad (9)$$

where $A^{-1}$ means the inverse matrix of A. It can be calculated by SVD method [3].

The design of the distribution of correctors is a significant part of closed orbit correction. To decrease the strength of correctors, according to formula (6), the correctors should be installed at the positions where β function is large. On the other hand, placing them at the sources of large field errors is conducive to more efficient correction. If there are k correctors located at such positions respectively, the response matrixes can be written as

$$A = \begin{pmatrix} a_{11} & \cdots & a_{1m} \\ \vdots & a_{ij} & \vdots \\ a_{n1} & \cdots & a_{nm} \end{pmatrix}$$
$$A_1 = \begin{pmatrix} a_{11} & \cdots & a_{1k} & 0 & \cdots & 0 \\ \vdots & \ddots & a_{ik} & \vdots & \ddots & \vdots \\ a_{n1} & \cdots & a_{nk} & 0 & \cdots & 0 \end{pmatrix}_{n\times m} \quad (10)$$

The correction process is divided into two parts. First, these k correctors offset the large field errors. Then they do closed orbit correction with the others. Supposing that the offset vector is $(b_k)$ and the COD aroused by it is $(p_{i1})$, the correcting vector is $(b_{j1})$, and the whole correcting vector is $(b_j)$, the process may be expressed as

$$(p_i) = -A \cdot (b_j), \text{ and } (p_{i1}) = A_1 \cdot (b_k). \quad (11)$$

$$(p_{i1} + p_i) = (p_i) + A_1 \cdot (b_k) = -A \cdot (b_{j1}). \quad (12)$$

The sum of vectors representing two steps is

$$\begin{aligned}(b_k + b_{j1}) &= (b_k) - A^{-1}[(p_i) + A_1 \cdot (b_k)] \\ &= (b_j) + (b_k) - A^{-1} A_1 \cdot (b_k)\end{aligned}. \quad (13)$$

The difference between $A^{-1}A_1$ and unit matrix I is that all elements are zero after $k^{th}$ row in $A^{-1}A_1$ since $A^{-1}A = I$. As the elements of $(b_k)$ are zero too after $k^{th}$ row, there is relationship

$$(b_k + b_{j1}) = (b_j). \quad (14)$$

From the above we may find that such arrangement can offset large error sources.

## 4 The closed orbit correction of HIMM

Among the 8 BPMs of HIMM synchrotron, 6 are located at the peaks of $\beta_y$ function, corresponding to the bottoms of $\beta_x$. Only 2 of them are located at the nearly maximum peaks of $\beta_x$. It means the correction in vertical plane will be better than that in horizontal one. The β functions are show in Fig. 1.

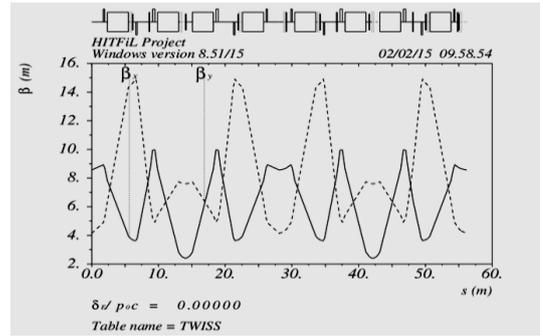

Fig. 1.  β functions of HIMM synchrotron

The alignment errors of the synchrotron are shown in Table 1.

Table 1.  Random alignment errors of HIMM (rms)

| Magnet | Δx, Δy/mm | Δs/mm | Δφ, Δθ/ mrad | Δψ/ mrad |
|---|---|---|---|---|
| Dipole | 0.5 | 1 | 0.5 | 0.2 |
| Quadrupole | 0.1 | 0.5 | 0.5 | 0.5 |

Under the linear condition, only the errors of dipole and quadrupole magnets are considered. With the maximum field errors of the magnets taken into account, the results of the simulations are shown in Fig. 2 - 5.

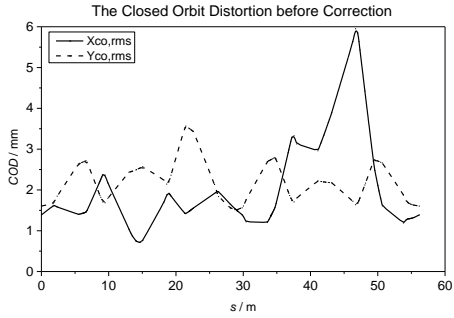

Fig. 2. Before correction, the COD at every point (the rms value of 30 simulations with different random numbers). It shows the distribution of the COD in the whole ring.

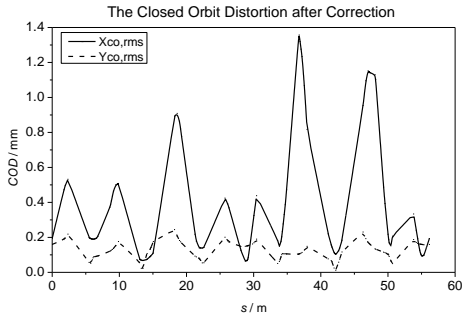

Fig. 3. After correction, the COD at every point.

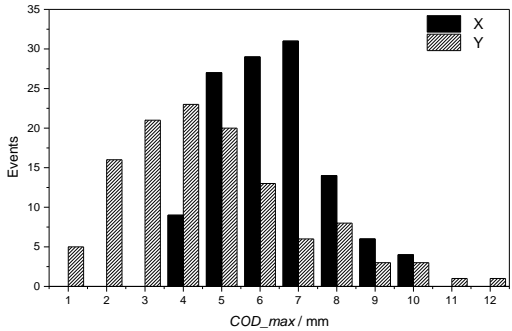

Fig. 4. Before correction, random distribution of the maximum COD of the whole ring. There are 120 times simulations with different random numbers.

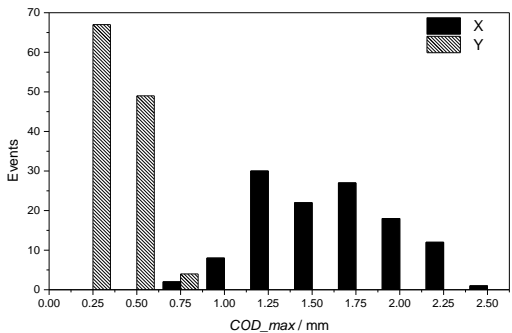

Fig. 5. After correction, random distribution of the maximum COD of the whole ring.

It is found that the variations of COD and β function have the same tendency. The maximum CODs after correction in vertical direction (Y) are lower than 0.5 mm in general. And in horizontal direction (X), the average value of the maximum CODs is nearly 1.6 mm. The CODs in two transverse directions are well corrected. But in X direction, it is a bit larger.

## 5  Analyses of the results

The distribution of BPMs and correctors in horizontal plane leads to the imperfection of correction. Another important reason affecting the effect of correction in horizontal direction is the alignment error in longitudinal direction ($\Delta s$). While the standard deviation of the random value of $\Delta s$ decreases, the closed orbit before/after correction is optimized. Mechanism of the change of the closed orbit arising from $\Delta s$ is shown in Fig. 6.

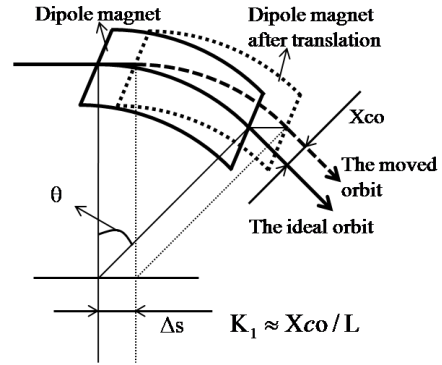

Fig. 6. The change of Xco (COD in X direction) arising from $\Delta s$. The arc of the sector on the left (the solid one) is the ideal orbit in a dipole magnet, and the one on the right (the dotted one) is the translational orbit caused by the effect of $\Delta s$. $\theta$ is the deflection angle of the dipole magnet and L is its length.

As shown in Fig. 6, this kind of COD is different from the distortion caused by dipole field error. The growth of COD happens in dipole magnet alone. The result may be described as generation of a parallel moved orbit. Approximately the effect is the same as two kickers providing two inverse deflection angles ($\pm K_1$) respectively at the entrance and exit of the magnet. The relationship between the increase of COD and $\Delta s$ is

$$X_{co} = \Delta s \cdot \sin\theta. \quad (15)$$

When $\theta$ is tiny, the distortion can be ignored. While the $\theta$ is large enough, we need two kickers to correct the distortion simultaneously. The traditional correction method can't decrease this kind of COD efficiently.

The correction results of different Δs random errors (respectively 0.5 mm, 1 mm, and 2 mm) are show in Fig. 7.

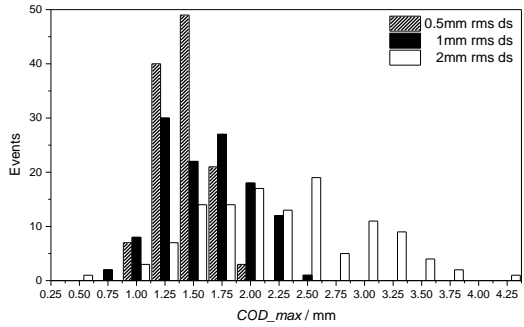

Fig. 7. After correction, random distribution of the maximum COD. There are 120 times simulations with different random numbers.

The reduction of the average value of the maximum CODs after correction nearly meets formula (15) when Δs are not too small.

## 6 Summaries

The main accelerator of HIMM is so compact that the number and azimuths of BPMs are hard to change. The locations of correctors are limited as well. The present result of closed orbit correction has already met the requirement. One significant measure to do further optimization of the COD is to improve the precision of the longitudinal alignment of dipole magnet.

## References

[1] Qin Q, High Energy Circular Accelerator Physics, Institute of High Energy Physics.
[2] J. Pierre, Trajectory and Closed Orbit Correction, CERN, CH-1211, Geneva 23.
[3] K. Baker, Singular Value Decomposition Tutorial, March 29, 2005.
[4] E. D. Courant, H. S. Snyder, Theory of the Alternating Gradient Synchrotron, Brookhaven National Laboratory, July15, 1957.
[5] A. Valentinov, I. Krylov, I. Yupinov, COD Measuring and Correction at SIBERIA-2, RRC Kurchatov Institute, Moscow, Russia, Proceeding of RuPAC XIX, Dubna 2004.
[6] B. Autin, Y. Marti, Closed Orbit Correction of A.G. Machines Using a Small Number of Magnets, Geneva, Switzerland, March 20, 1973
[7] N. Nakamuraa, H. Takakia, et al. New orbit correction method uniting global and local orbit corrections, Nuclear Instruments and Methods in Physics Research A, 2006.